\documentclass[12pt]{article}
\usepackage[dvips]{color,graphicx}
\usepackage{hyperref}
\usepackage{amsmath,amssymb}
\DeclareGraphicsExtensions{.eps}

\newcommand{\ls}{\ensuremath{l_s}} 

\def\p{\partial}

\def\slash#1{\ensuremath{\;/\!\!\!\! #1}}

\newcommand{\cF}{{\mathcal{F}}}

\newcommand{\cL}{\mathcal{L}}

\newcommand{\cN}{{\mathcal{N}}}
\newcommand{\cO}{{\mathcal{O}}}

\newcommand{\bS}{{\mathbf{S}}}

\newcommand{\be}{\begin{equation}}
\newcommand{\ee}{\end{equation}}
\newcommand{\bea}{\begin{eqnarray}}
\newcommand{\eea}{\end{eqnarray}}

\begin{document}

\begin{titlepage}

\begin{flushright}
UTTG-05-11
\end{flushright}

\begin{center} \Large \bf Early-Time Energy Loss in a \\
Strongly-Coupled
 SYM Plasma
\end{center}

\begin{center}
Alberto G\"uijosa$^{\dagger}$\footnote{alberto@nucleares.unam.mx}
and Juan F.~Pedraza$^{\star}$\footnote{jpedraza@physics.utexas.edu}

\vspace{0.2cm}
$^{\dagger}\,$Departamento de F\'{\i}sica de Altas Energ\'{\i}as, Instituto de Ciencias Nucleares, \\ Universidad Nacional Aut\'onoma de
M\'exico,\\ Apartado Postal 70-543, M\'exico D.F. 04510, M\'exico\\
 \vspace{0.2cm}
$^{\star}\,$Theory Group, Department of Physics, University of Texas,\\
1 University Station C1608, Austin, Texas 78712, USA\\
\vspace{0.2cm}
\end{center}

\begin{center}
{\bf Abstract}
\end{center}
\noindent
We carry out an analytic study of the early-time motion of a quark in a strongly-coupled maximally-supersymmetric Yang-Mills plasma, using the AdS/CFT correspondence. Our approach extracts the first thermal effects as a small perturbation of the known quark dynamics in vacuum, using a double expansion that is valid for early times and for (moderately) ultrarelativistic quark velocities. The quark is found to lose energy at a rate that differs significantly from the previously derived stationary/late-time result: it scales like $T^4$ instead of $T^2$, and is associated with a friction coefficient that is not independent of the quark momentum. Under conditions representative of the quark-gluon plasma as obtained at RHIC, the early energy loss rate is a few times smaller than its late-time counterpart. Our analysis additionally leads to thermally-corrected expressions for the intrinsic energy and momentum of the quark, in which the  previously discovered limiting velocity of the quark is found to appear naturally.

\vspace{0.2in}
\smallskip
\end{titlepage}
\setcounter{footnote}{0}

\tableofcontents

\section{Introduction and Summary}

\subsection{Background: Energy Loss and AdS/CFT}

Heavy ion collision experiments at RHIC and, recently, LHC, have amassed substantial evidence supporting the detection of the long-sought quark-gluon plasma (QGP), a hot and dense phase of deconfined strongly-interacting matter \cite{qgprev}. Energetic partons serve as important probes of this thermal medium, and there exists an enormous `jet-quenching' literature dedicated to analyzing the manner in which the plasma damps their motion, and is in turn disturbed by their passage \cite{energylossrev}. Given the experimental indications that the QGP is strongly-coupled, perturbative QCD is believed to be inadequate for at least some aspects of the relevant calculations, creating a demand for new theoretical tools. In the past five years, interesting steps have been taken towards meeting this demand via the AdS/CFT, or, more generally, gauge/gravity correspondence \cite{malda,magoo}, starting with the seminal works \cite{hkkky,gubser,ct,liuqhat} and continuing with a large body of work that has been reviewed in \cite{clmrw,hr,esz,gk,mateos,pz}.

The gauge/gravity correspondence is by now well-established as a tool that grants access (often analytically) to a large class of strongly-coupled non-Abelian gauge theories, via a drastic and surprising rewriting in terms of string-theoretic (frequently just supergravity) degrees of freedom living on a curved higher-dimensional geometry.    The gauge/gravity catalog includes some theories that display characteristic QCD-like physics such as confinement and chiral-symmetry breaking, but not, to date, QCD itself. Nonetheless, the correspondence has already been very successful in developing our intuition on the behavior of strongly-coupled gauge theories, even to the point of providing useful suggestions for phenomenological models of the QGP \cite{horowitz,hg,mr,kharzeev}.

In the gauge/gravity context, a strongly-coupled thermal gluon plasma is described by a black hole geometry with special asymptotics, and a quark traversing the plasma is dual to a string moving on this background. More precisely, the quark corresponds to the tip of a string, whose
body codifies the profile of the non-Abelian (near and radiation) fields sourced by the quark. As the string endpoint moves, its body lags behind it, exerting a drag on the tip that is the gravity-side realization of the damping force exerted by the plasma on the quark, as studied in \cite{hkkky,gubser}.\footnote{It was later shown in \cite{lorentzdirac,damping} that this trailing string mechanism is in fact much more general, and accounts even for the radiative damping expected in vacuum. That is, irrespective
of whether a \emph{spacetime} black hole is present or not, the body of the string plays the role of an energy sink, as befits its identification as the embodiment of the gluonic degrees of freedom. On the other hand, energy loss via the string does turn out to be closely
associated with the appearance of a \emph{worldsheet} horizon,
  as noticed initially in \cite{gubserqhat,ctqhat} at finite temperature and in
  \cite{dragtime} (see also \cite{dominguez,xiao}) for the zero temperature case.}
These works focused on the particular case of a plasma in maximally-supersymmetric ($\mathcal{N}=4$) Yang-Mills theory (MSYM), which is the simplest and best-understood example of the correspondence, and is known to furnish a useful (albeit rudimentary) toy model of the QGP. MSYM is a conformally-invariant theory (CFT), whose dual gravity description at finite temperature involves a black hole that is asymptotically anti-de Sitter (AdS). For concreteness, we will hereafter concentrate on this same AdS/CFT setup, even though we expect most of the qualitative lessons of our work to be more general.\footnote{Energy loss studies have been carried out via the gauge/gravity correspondence in many other theories, including some with properties more akin to those of QCD--- see, e.g., \cite{otherdragforce,hk} and references therein. Other examples of drag force calculations may be found in the reviews \cite{clmrw,hr,esz,gk,mateos,pz}.}

The analysis of \cite{hkkky,gubser} led to explicit formulas for the rate of energy and momentum radiated by a heavy quark as it moves through an infinite static MSYM plasma,
\begin{equation}\label{elossgubser}
\frac{dE_{\mbox{\scriptsize rad}}}{dt}=\frac{\pi}{2}\sqrt{\lambda}T^2
\frac{v^2}{\sqrt{1-v^2}}
\end{equation}
and
\begin{equation}\label{plossgubser}
\frac{dP_{\mbox{\scriptsize rad}}}{dt}=
\frac{\pi}{2}\sqrt{\lambda}T^2\frac{v}{\sqrt{1-v^2}}~,
\end{equation}
where $\lambda$ denotes the 't~Hooft coupling.\footnote{Various works have explored the way in which these heavy quark results are modified for heavy sources of the gluonic field in color representations other than the fundamental (including the adjoint) \cite{draggluon}, as well as for \emph{light} quarks and gluons \cite{lightpartonenergyloss}.}
In the context of weakly-coupled QCD, it is known that energy loss at high parton velocity is dominated by the strong-interaction analog of bremsstrahlung, i.e., medium-induced gluonic radiation, while for low velocities collisional loss becomes important \cite{energylossrev}. In the strongly-coupled MSYM setup made available to us by AdS/CFT, the perturbative terminology is no longer adequate. The flow of energy/momentum along the body of the trailing string corresponds to energy/momentum transported away from the quark via the gluonic field generated jointly by the quark and the plasma, in a pattern that has been meticulously studied in a large body of work that began with \cite{gluonicprofile}, has been reviewed in \cite{gluonicprofilerev} and includes the interesting recent additions \cite{gluonicprofilerecent}. For ease of language, throughout this paper we will continue to speak of radiation, even though this concept is not really appropriate within an infinite thermal medium.

We should also stress that, within the framework of \cite{hkkky,gubser}, adopted also in this paper, the entire energy loss calculation is carried out inside the MSYM theory, whose coupling does not run and is taken to be large. For quarks that are extremely energetic, the asymptotic freedom of QCD would lead one to expect deviations between this scenario and the real-world QGP, and seek a hybrid approach describing hard/perturbative emission of a gluon that subsequently propagates through a strongly-coupled medium, as advocated in \cite{liuqhat,liuwilson,mr}.\footnote{The approach of \cite{liuqhat,liuwilson} involves the extraction from a lightlike Wilson loop of a transport (`jet-quenching') parameter $\hat{q}$, which characterizes the medium and whose value at strong coupling curiously differs from the customary definition of $\hat{q}$ as the average transverse momentum picked up per unit distance traveled \cite{gubserqhat,ctqhat}. Some doubts about the procedure proposed in \cite{liuqhat,liuwilson} to calculate $\hat{q}$ on the gravity side of the correspondence had been raised by, e.g., \cite{dragqqbar,argyres2,argyres3}, but these concerns appear to have been resolved by a recent correction to the procedure \cite{elr}, which happens to leave the result unchanged.} It is debatable, however, which of these two scenarios is more appropriate at the not extremely relativistic heavy quark energies achieved in RHIC. At the very least, our calculations have a direct interpretation in terms of energy loss in the strongly-coupled MSYM plasma, which is an interesting theoretical question in its own right.

\subsection{Motivation, outline and main results}

Naturally, expressions  (\ref{elossgubser}) and (\ref{plossgubser}) were derived under a number of simplifying assumptions. The one that matters most for our purposes is the restriction made by the authors of \cite{hkkky,gubser} to a configuration where the quark is either moving with constant (possibly relativistic) velocity as a result of being pulled by an external force that precisely balances the drag, or is unforced but moving nonrelativistically and about to come to rest. Either of these scenarios requires the interaction between the quark and the medium to occur over a considerable period of time.  Since the actual QGP produced at RHIC or LHC has a finite temporal and spatial extent, it is questionable whether a heavy quark propagating through it will have enough time to reach a quasi-stationary configuration, or to be nearly stopped (in which case it would anyhow not lead to an experimentally distinguishable jet).

The actual energy/momentum loss
  might thus be expected to differ from (\ref{elossgubser}) and (\ref{plossgubser}) in a situation where the
  quark moving through the plasma is accelerating, or in the initial
  period  following its production within the thermal medium.
  This point has been emphasized from the phenomenological perspective
  in \cite{peigne1,peigne2}.
 The estimates there are based on perturbative
  calculations, so it is interesting to inquire into this issue in
  the strongly-coupled systems available to us through
  the AdS/CFT correspondence.

  Now,
  the restriction in  \cite{hkkky,gubser}
  to the stationary or asymptotic cases
   was of course implemented
  to gain analytic control on the problem of energy loss.
  The authors of \cite{beuf} were able to slightly extend the stationary result to the case of a slowly decelerating quark, working in an expansion in powers of $\sqrt{\lambda}T/m$.
  Away from these regimes, it is difficult to follow
  the evolution of the quark in
  the thermal plasma, or equivalently, of the string
  on the AdS black hole geometry.
  One possibility is to resort
  to numerical analysis, a strategy adopted in \cite{dragtime}.
 Among other things, that work studied
  the case where a quark that is initially static within the plasma is accelerated by an
    external force over a finite period of time and is thereafter
    released. It was found that
    under such conditions,
    and for values of the mass in the neighborhood of the charm quark,
    there exists a period after release where the quark
    dissipates energy at a rate that is substantially \emph{smaller} than
    the stationary/asymptotic result (\ref{elossgubser})
    obtained in \cite{hkkky,gubser}. In addition,
    the rate of energy loss in the initial stage where the quark is
    externally forced was found to be almost completely accounted for
    by a generalized Lienard formula describing radiation in vacuum (derived in that same paper and in \cite{lorentzdirac,damping}).
    Unfortunately, the numerical integration
degraded rather quickly, as a result of which the investigation of \cite{dragtime} was
limited to intervals that are more than an order of magnitude below the
experimental timescale of the plasma (typically
$t_{\mbox{\scriptsize breakdown}}\sim 0.9 /\pi T\sim
0.3~\mbox{fm}/c$), and to rather small quark velocities
at the time of release (for the most part, $v_{\mbox{\scriptsize
release}}<0.1$).

Both to try to overcome these numerical limitations, and for the additional intuition it would provide, it is clearly desirable to gain some measure of analytic control over the problem of early-time quark damping in a strongly-coupled plasma, which is what we set out to accomplish in the present paper. Our approach will be to study thermal effects as a small perturbation on the zero-temperature evolution, which is reasonable on general physical grounds during the initial stage of motion through the plasma, and is moreover supported by the numerical results of \cite{dragtime}. For simplicity, we will restrict attention to the case where the quark moves solely along one direction, denoted by $x$, as this is enough to examine the damping we are after.

After setting up our language and notation in Section \ref{quarksec} (which can be safely skipped by the cognoscenti), we begin to develop our story by establishing the zero temperature framework over which we will perturb. In Section \ref{embeddingsubsec} we recall the string embedding (\ref{mikhsol}), argued to be an extremum of the Nambu-Goto action in \cite{mikhailov}, and explicitly verify that it indeed solves the corresponding nonlinear equation of motion. This provides the string configuration relevant for an \emph{arbitrary} timelike motion of the heavy quark in vacuum, under the condition that waves on the string be purely retarded, to capture the gluonic fields causally set up by the quark. In Section \ref{dampingsubsec} we then present a streamlined version of the procedure of \cite{dragtime,lorentzdirac,damping} that uses the string profile (\ref{mikhsol}) to turn the standard string boundary condition (\ref{stringbc}) into a dynamical equation for the quark in vacuum. The result, equation (\ref{peom}), is a generalized version of the classic Lorentz-Dirac equation \cite{lorentzdirac,damping} that is physically sensible, and incorporates expressions for the intrinsic momentum of the quark (\ref{pq}) and the rate (\ref{prad}) at which it radiates momentum. Equivalently, equation (\ref{eom}) incorporates expressions for the intrinsic quark energy (\ref{eq}) and rate of energy radiation (\ref{erad}). As found in \cite{dragtime,lorentzdirac,damping}, for a quark with finite mass and therefore finite size, these expressions differ from the naive ones by corrections that depend on the external force and the quark Compton wavelength.
In the quark equation of motion we can solve for the acceleration to arrive at (\ref{alinear}), which shows that, as expected (and unlike in the standard Lorentz-Dirac equation), an unforced quark in the vacuum of MSYM must move at constant velocity. This is of course the property that we expect to be modified upon the introduction of a thermal medium.

In Section \ref{embeddingtempsubsec}, the finite-temperature string equation of motion (\ref{ngeomtemp}) is considered for early times in the evolution, when the disturbed portion of the string remains far from the black hole horizon, as indicated in (\ref{smalltemp}), and with the embedding (\ref{yexpansion}) deviating only slightly from its zero-temperature counterpart (\ref{mikhsol}), to arrive at  the linearized equation (\ref{linearngeom}). This still contains a very complicated dependence on the quark trajectory, so, motivated by the experimental scenario, we further restrict attention to quarks traveling rapidly through the plasma. In short, we work in the early-time and moderately ultrarelativistic regime (\ref{smallparameters}), which essentially amounts to
\begin{equation}\label{smallparameterssimp}
\pi^4 T^4 d^4 \ll 1-v^2\ll 1~,
\end{equation}
where $d\sim t/\gamma$
is the size of the growing region of the gluonic field that has been disturbed by the motion of the quark.
Under these conditions we are ultimately led to the thermally-corrected embedding (\ref{xcorrected2}).

We then go on in Section \ref{dampingtempsubsec} to deduce the way in which the first thermal correction to the embedding affects the evolution of the quark, focusing attention on the case of prime phenomenological relevance where the quark is not externally forced. By imitating the procedure used in the zero-temperature setting, we derive an equation of motion for the quark in the thermal medium.  Along the way, we find that our framework naturally makes contact in (\ref{vvtildesimpl}) with the previously known limiting velocity $v_m<1$ (defined in (\ref{vm})) for a quark immersed in the strongly-coupled plasma \cite{argyres1,gubserqhat,ctqhat,mateosvm,liu4,argyres3,dragtime,ms,st}.
The thermal equation of motion that results from our analysis can be written in either of the alternative forms (\ref{atemp}), (\ref{esplit}) or (\ref{psplit}), with the identification of (\ref{eqtemp}) and (\ref{pqtemp}) as the thermally corrected intrinsic energy and momentum of the quark (which are seen to diverge as $v\to v_m$), and of
\begin{equation}\label{eradtempintro}
 {d E_{\mbox{\scriptsize rad}}\over dt}
 =\frac{\pi}{16}\frac{\lambda^{3/2}T^4}{m^2}\frac{v/v_m}{1-(v/v_m)^2}
 \end{equation}
 and
\begin{equation}\label{pradtempintro}
 {d P_{\mbox{\scriptsize rad}}\over dt}
 =\frac{\pi}{16}\frac{\lambda^{3/2}T^4}{m^2}\frac{1}{1-(v/v_m)^2}~
 \end{equation}
 as the rates at which the quark loses energy and momentum. These are then our main results.

 Section \ref{phenosec} discusses the possible phenomenological implications of our findings. After verifying that our approximations are well-suited to the typical energetic heavy quarks obtained at RHIC, we contrast our early-time results (\ref{eradtempintro})-(\ref{pradtempintro}) against the late-time expressions
 (\ref{elossgubser})-(\ref{plossgubser}) obtained in \cite{hkkky,gubser}. In (\ref{latevsearly}) we see that, for parameter values appropriate to RHIC, the former can be up to six times \emph{smaller} than the latter. This comparison is consistent with the (more limited) numerical results of \cite{dragtime}. The functional form of our results is also of interest. {}From our derivation it becomes clear that the $T^4$-dependence seen in (\ref{eradtempintro})-(\ref{pradtempintro}) (and anticipated in an estimate based on saturation physics \cite{dominguez}) follows naturally from the first thermal corrections in the metric, and via dimensional analysis, this fixes the accompanying dependence on the quark mass, which in turn explains the peculiar power of the 't~Hooft coupling. Another prominent feature of our formulas is their velocity-dependence: whereas the late-time friction coefficient (\ref{mulate}) deduced from (\ref{elossgubser})-(\ref{plossgubser}) is momentum-independent,
 its early-time counterpart (\ref{muearly}) depends strongly on momentum. This feature should constitute an interesting experimental signature when used as input for phenomenological models, as in \cite{horowitz,hg,mr,kharzeev}.

 As envisioned from the outset, our investigation is limited to the initial stage where the quark is only starting to feel the effects of the plasma. In (\ref{ttemp2}) we learn that our results should be reliable for times that are not too close to the lifetime of the real-world QGP.
 The temporal range of validity of our work is an improvement over the one accessible in the numerical work of \cite{dragtime}, but still suggests that the experimental results should effectively arise from some sort of average or interpolation between the early and late energy loss rates (\ref{eradtempintro}) and
 (\ref{elossgubser}).
 In this connection it should also be kept in mind that a separate numerical study in \cite{dragtime} (building on \cite{hkkky}) of a quark that is created within the plasma back-to-back with its corresponding antiquark found the late-time energy loss rates of \cite{hkkky,gubser} to be relevant essentially as soon as the quark and antiquark separated beyond the velocity-dependent screening length determined independently in \cite{liuwind,dragqqbar} (see also \cite{sonnenschein}) and \cite{dragtime} itself.
 The way in which these findings interface with ours is also discussed in Section \ref{phenosec}. The upshot is that, in the more realistic situation where the presence of the accompanying antiquark is taken into consideration, our results would be most relevant in the stage before the separating quark and antiquark lose contact with one another. This is natural, because screening emerges only when thermal effects are substantial. For a quark ploughing through a significant portion of the plasma, then, the late-time formulas (\ref{elossgubser})-(\ref{plossgubser}) actually stand a chance of controlling the larger portion of the evolution.

Although we have succeeded in gaining some analytic control on parton energy loss in a plasma within a novel temporal regime, clearly more work will be needed to shed additional light on the overall problem and to remove the various idealizations adopted in the existing calculations. Based on past experience, it seems likely that the gauge/gravity correspondence will continue to be a useful guide into the strong-coupling aspects of this problem.

\section{Preliminaries: Quarks as Strings}\label{quarksec}

 Maximally supersymmetric (i.e., $\cN=4$) Yang-Mills (MSYM) is a  conformally invariant theory comprising a gauge field, 6 real scalar fields and 4 Weyl fermions, all in the adjoint representation of the gauge group.
According to the AdS/CFT correspondence \cite{malda}, large-$N_c$ strongly-coupled $SU(N_c)$ MSYM on $(3+1)$-dimensional Minkowski spacetime,
with 't Hooft coupling $\lambda\equiv g_{YM}^2 N_c$ and temperature $T$, is equivalent to Type IIB string theory living  on the
(planar Schwarzschild-AdS)$_5\times\bS^5$ geometry
\begin{eqnarray}\label{metric}
ds^2&=&G_{mn}dx^m dx^n={R^2\over z^2}\left(
-hdt^2+d\vec{x}^2+{dz^2 \over h}\right)+R^2 d\Omega_5^2~, \\
h&=&1-\frac{z^{4}}{z_h^4}~, \qquad {R^4\over \ls^4}=\lambda~, \qquad
z_h={1\over \pi T}~,\nonumber
\end{eqnarray}
(with a constant dilaton and $N_c$ units of Ramond-Ramond
five-form flux through the five-sphere,) where $\ls$ denotes the string length.
The
radial direction $z$ is mapped holographically into a variable length scale in the gauge theory
\cite{uvir}. The coordinates $x^{\mu}\equiv(t,\vec{x})$ are parallel to the AdS boundary $z=0$ and are
directly identified with the gauge theory spacetime coordinates, while the five-sphere coordinates are associated with the global $SU(4)$ internal (R-) symmetry of MSYM.

The state of IIB string theory described by the pure AdS$_5\times\bS^5$ geometry
given by the $z_h\to\infty$ limit of (\ref{metric}) corresponds to the vacuum of the MSYM theory, and the closed
string sector describing (small or large) fluctuations on top of it fully captures the gluonic ($+$
\emph{adjoint} scalar and fermionic) physics. To this closed string sector belongs in particular the $z_h<\infty$ geometry (\ref{metric}), which describes a black hole (strictly, black brane)  with an event horizon at $z=z_h$ and Hawking temperature $T$, and is dual to a MSYM thermal ensemble at the same temperature. This then is the string theory implementation of our desired plasma of gluons ($+$ scalars and fermions).

To this thermal MSYM setup we can add $N_f$ hypermultiplets
(each composed of a Dirac fermion and 2 complex scalars)
in the \emph{fundamental}
representation of the $SU(N_c)$ gauge group, breaking the supersymmetry down to $\cN=2$.
These are the degrees of freedom that we will refer to as `quarks,' even though they include both spin
$1/2$ and spin $0$ fields. In string theory language, adding these hypermultiplets corresponds \cite{kk}
to the introduction of an open string sector associated with a
stack of $N_f$ D7-branes in the geometry
(\ref{metric}). For $N_f\ll N_c$, the backreaction of the D7-branes on the geometry can
be neglected; in the gauge theory this corresponds to working in a `quenched' approximation
that ignores quark loops (as well as the positive beta function they would generate).

The D7-branes
cover the four gauge theory directions $t,\vec{x}$, and extend
along the radial AdS direction up from the boundary at $z=0$ to a
position where they `end' (meaning that the $\bS^3\subset\bS^5$
that they are wrapped on shrinks down to zero size), whose
location $z=z_m$ is related to
the mass $m$ of the
quark through \cite{hkkky}
\begin{equation}\label{zm} {1\over z_m}={2\pi
m\over\sqrt{\lambda}}\left[1+{1\over 8}\left(\sqrt{\lambda} T\over 2 m\right)^4-{5\over
128}\left(\sqrt{\lambda} T\over 2 m\right)^8+\cO\left(\left(\sqrt{\lambda} T\over 2
m\right)^{12}\right)\right]~.
\end{equation}
We are interested in quarks that are heavy in the sense that $m\gg \sqrt{\lambda}T/2$, so we will restrict attention to D7-brane embeddings with $z_m\ll z_h$, and in so doing will only need the leading order approximation to (\ref{zm}),
\begin{equation}\label{zmnoplasma}
z_m={\sqrt{\lambda}\over 2\pi m}~.
\end{equation}

An isolated quark is dual to an open string that extends radially from the D7-branes to the black hole
horizon at $z=z_h$. We take the string
to lie, consistently
with its equations of motion, at the `North Pole' on the $\bS^5$ (the point where the
$\bS^3\subset\bS^5$ that the D7-branes are wrapped on collapses to zero size), so the angular
components of the metric will not play any role in our work. This is natural for eventual application to the real-world QGP, because QCD has no symmetry on a par with the $SU(4)$ internal symmetry of MSYM that is dual to the five-sphere isometry group. The lower endpoint of our string will then
necessarily be located at $z=z_m$.

As is customary in string theory, we will describe the dynamics
 of our string in first-quantized language, and since we take it to be heavy, we are allowed to treat it semiclassically. In gauge theory language,
then, we are coupling a first-quantized quark to the gluonic ($+$ other MSYM) field(s), and
carrying out the full thermal path integral over the strongly-coupled field(s) (the result of which is
codified by the Schwarzschild-AdS spacetime), but treating the path integral over the quark trajectory
$\vec{x}(t)$ in a saddle-point approximation.\footnote{Starting with \cite{rangamani,sonteaney}, various works have explored quantum fluctuations of the quark about certain types of
average trajectories--- see, e.g., \cite{brownian,gursoy,cp} and references therein. These fluctuations are suppressed by a factor of $1/\sqrt{\lambda}$, and would have to be complemented with contributions from the path integral of the type explored in a somewhat different context in \cite{fluct}.}

It is useful to keep in mind that in the nonperturbative context provided to us by the AdS/CFT correspondence, a quark with finite mass ($z_m>0$) is automatically not `bare' but `composite' or `dressed'.  This can be inferred, for instance, from the expectation value of the gluonic field surrounding a static quark
\cite{martinfsq}, or from the deformed nature of the quark's dispersion relation \cite{dragtime,lorentzdirac,damping}.
As seen in those calculations, the characteristic thickness of the `gluonic cloud' surrounding the quark
 is given by $z_m$, which is thus understood to be the analog of the Compton wavelength
for our non-Abelian source.

The string dynamics follows as usual from the Nambu-Goto action
\begin{equation}\label{nambugoto}
S_{\mbox{\scriptsize NG}}=-{1\over 2\pi\ls^2}\int
d^2\sigma\,\sqrt{-\det{g_{ab}}}\equiv {R^2\over 2\pi l_s^2}\int
d^2\sigma\,\cL_{\mbox{\scriptsize NG}}~,
\end{equation}
where $g_{ab}\equiv\p_a X^m\p_b X^n G_{mn}(X)$ ($a,b=0,1$) denotes
the induced metric on the worldsheet. For convenience we choose to work in the static gauge
$\sigma^0=t$, $\sigma^1=z$.

By the AdS/CFT dictionary, it is the endpoint of the string on the D7-branes that directly corresponds to the quark, and so the latter's trajectory can be read off as
\begin{equation}\label{quarkx}
\vec{x}(t)=\vec{X}(t,z_m)~.
\end{equation}
In our analysis below we will have need to refer to
the velocity, acceleration and jerk of the quark, which will be denoted as $\vec{v}\equiv d\vec{x}/dt$,
$\vec{a}\equiv d\vec{v}/dt$, $\vec{j}\equiv d\vec{a}/dt$.
Since the damping force we are after will be
exerted by the plasma against the quark's direction of motion, for simplicity we will consider
motion and deformation of the
string purely along a single direction $x\equiv x^1$. Extension of our entire study
to arbitrary three-dimensional motions does not require any new conceptual ingredients (and at zero temperature has already been carried out in \cite{lorentzdirac,damping}),
 but is algebraically much more complicated (especially since at finite-temperature one can no longer rely on Lorentz invariance as in \cite{lorentzdirac,damping}).

Under the simplifying conditions described in the two previous paragraphs,
the Nambu-Goto Lagrangian boils down to
\begin{equation}\label{nambugotosimpl}
\cL_{\mbox{\scriptsize NG}}=-\sqrt{1+h {X^{'}}^{2}-{\dot{X}^2\over h}}~,
\end{equation}
and the non-zero
canonical momentum densities
$\Pi^a_{\mu}\equiv\p\cL_{\mbox{\scriptsize NG}}/\p(\p_a X^{\mu})$
are given by
\begin{eqnarray}\label{momenta}
\Pi^t_t&=&-\frac{h{X^{'}}^{2}+1}{z^2\sqrt{1+h {X^{'}}^{2}-{\dot{X}^2\over h}}}~,\nonumber\\
\Pi^t_x&=&\frac{\dot{X}}{z^2 h\sqrt{1+h{X^{'}}^{2}-{\dot{X}^2\over h}}}~,\\
\Pi^z_t&=&\frac{h\dot{X}X'}{z^2\sqrt{1+h{X^{'}}^{2}-{\dot{X}^2\over h}}}~,\nonumber\\
\Pi^z_x&=&-\frac{h X'}{z^2\sqrt{1+h{X^{'}}^{2}-{\dot{X}^2\over
h}}}~,\nonumber
\end{eqnarray}
where of course $\dot{X}\equiv\p_t X(t,z)$, $X'\equiv\p_z X(t,z)$.
Notice that, due to our normalization of $\cL_{\mbox{\scriptsize
NG}}$, the $\Pi^a_{\mu}$ must be multiplied by
$R^2/2\pi\ls^2=\sqrt{\lambda}/ 2\pi$ to obtain the physical energy
and momentum densities.

We can exert an external force $\vec{F}$ on the quark/string endpoint by turning on an electric field
$F_{0i}=F_i$ on the D7-branes, which contributes to the string action the term
\begin{equation} \label{couplingtoA} S_{\mbox{\scriptsize F}}=\int
dt\,A_{\mu}(X(t,z_m))\p_{t}X^{\mu}(t,z_m)~. \nonumber
\end{equation}
Variation of the string action $S_{\mbox{\scriptsize NG}}+S_{\mbox{\scriptsize F}}$ implies the
 Nambu-Goto equations of motion (or, equivalently, current conservation conditions)
 \begin{equation}\label{ngeom}
 \p_t\Pi^{t}_{\mu}+\p_z\Pi^{z}_{\mu}=0
 \end{equation}
  for all interior points of the string, plus the standard
boundary condition
 \begin{equation}\label{stringbc}
\Pi^{z}_{\mu}(t)|_{z=z_m}= \frac{2\pi}{\sqrt{\lambda}}\cF_{\mu}(t)\quad\forall~t~,
\end{equation}
where $\cF_{\mu}=-F_{\nu\mu}\p_{t}x^{\nu} =(-\vec{F}\cdot\vec{v},\vec{F})$ is the
Lorentz four-force. In the following sections we will solve for the dynamics of this system to extract the thermal damping rate.

\section{Zero Temperature}\label{zerotempsec}

In this section we will examine the quark dynamics in the MSYM vacuum, to lay the groundwork for the finite-temperature case that is our real interest. In so doing, we will rederive a subset of the results of \cite{mikhailov,dragtime,lorentzdirac,damping} in a form that is better suited to our purposes.

\subsection{String embedding}\label{embeddingsubsec}

At zero temperature ($z_h\to\infty$) the background (\ref{metric}) reduces to  pure AdS, and the string equation motion (\ref{ngeom}) reads
\begin{equation}
\frac{\partial}{\partial t}\left(\frac{\dot{X}}{z^2\sqrt{1+{X'}^2-\dot{X}^2}}\right)-\frac{\partial}{\partial z}\left(\frac{X'}{z^2\sqrt{1+{X'}^2-\dot{X}^2}}\right)=0~,
\end{equation}
which can be reprocessed to
\begin{equation}\label{eom}
\ddot{X}-X''+\ddot{X}X'^2+X''\dot{X}^2-2\dot{X}X'\dot{X}'+\frac{2X'}{z}+\frac{2X'^3}{z}-\frac{2X'\dot{X}^2}{z}=0~.
\end{equation}

To understand the behavior of the quark in the MSYM vacuum, our task is then to solve equation (\ref{eom})
subject to the forcing boundary condition (\ref{stringbc}), and read off the quark trajectory from the result using (\ref{quarkx}). Equivalently, we can take a specific quark trajectory and interpret it through (\ref{quarkx}) as a boundary condition for the string, solve for the corresponding string embedding, and in the end read off from (\ref{stringbc}) the forced required to guide the quark along the chosen trajectory. For the time being we will adopt this second perspective and, moreover, will choose to parametrize our desired string embedding in terms of the data for a quark that is infinitely massive, thus setting $z_m=0$ until further notice. In \cite{mikhailov}, Mikhailov used the symmetries of the AdS geometry to argue that, for an \emph{arbitrary} time-like quark trajectory $x(t)$, the embedding
\begin{eqnarray}\label{mikhsol}
X(t_r,z)&=&x(t_r)+\frac{v(t_r) z}{\sqrt{1-v(t_r)^2}}~,\\
t(t_r,z)&=&t_r+\frac{z}{\sqrt{1-v(t_r)^2}}~, \nonumber
\end{eqnarray}
which manifestly meets the condition (\ref{quarkx}), also extremizes the Nambu-Goto action, and must therefore be the solution we are seeking.

 {}From the structure of (\ref{mikhsol}) we see that the behavior of the string at a given time $t$ and radial depth $z$ is completely determined by the behavior of the quark/string endpoint at the earlier, \emph{retarded} time $t_r$. The definition of $t_r$ implicit in (\ref{mikhsol}) can be shown to follow from projecting back to the AdS boundary along a curve that is null on the string worldsheet \cite{mikhailov}, in analogy with the  Lienard-Wiechert story in classical electrodynamics. The embedding (\ref{mikhsol}) thus describes a wave on the string that is  \emph{purely outgoing}: it is generated at the string endpoint and then ascends along the body of the string, moving into the AdS bulk. This is why, among the infinite number of extremal string embeddings that satisfy (\ref{quarkx}), the profile (\ref{mikhsol}) is  the one that is of immediate physical interest for us: it is dual to a MSYM configuration where waves in the gluonic field move out from the quark to infinity.

 We will now show explicitly that (\ref{mikhsol}) indeed solves the equation of motion (\ref{eom}). Noting that
 \begin{eqnarray}\label{differentials}
dt&=&dt_r\left[{va z\over(1-v^2)^{3/2}}+1\right]+{dz\over\sqrt{1-v^2}}~,\\
dX&=&dt_r\left[{a z\over\sqrt{1-v^2}}+{v^2 a
z\over(1-v^2)^{3/2}}+v\right]+{v dz\over\sqrt{1-v^2}}~,\nonumber
\end{eqnarray}
where from now on it is understood that the velocity, acceleration and jerk of the quark (in the notation introduced after (\ref{quarkx})) are evaluated at $t_r$, we can deduce after some algebra that
\begin{equation}\label{xdot}
\dot{X}\equiv\left(\frac{\partial X}{\partial t}\right)_z=\left(\frac{\partial t_r}{\partial t}\right)_z\left(\frac{\partial X}{\partial t_r}\right)_z=\frac{az+v(1-v^2)^{3/2}}{vaz+(1-v^2)^{3/2}}~,
\end{equation}
\begin{equation}\label{xprime}
X'\equiv\left(\frac{\partial X}{\partial z}\right)_t=\left(\frac{\partial X}{\partial z}\right)_{t_r}-\left(\frac{\partial t}{\partial z}\right)_{t_r}\dot{X}=-\frac{az(1-v^2)^{1/2}}{vaz+(1-v^2)^{3/2}}~,
\end{equation}
and similarly
\begin{eqnarray}\label{seconderivatives}
\ddot{X}&\equiv&\left(\frac{\partial \dot{X}}{\partial t}\right)_z=\frac{-a^3z^2(1-v^2)^{3/2}+3va^2z(1-v^2)^{3}+jz(1-v^2)^{4}+a(1-v^2)^{9/2}}{[vaz+(1-v^2)^{3/2}]^3}~,
\nonumber\\
X''&\equiv&\left(\frac{\partial X'}{\partial z}\right)_t=
\frac{-a^3z^2(1-v^2)^{1/2}+va^2z(1-v^2)^{2}+jz(1-v^2)^{3}-a(1-v^2)^{7/2}}{[vaz+(1-v^2)^{3/2}]^3}~,
\nonumber\\
\dot{X}'&\equiv&\left(\frac{\partial X'}{\partial t}\right)_z=\frac{a^3z^2(1-v^2)-2va^2z(1-v^2)^{5/2}-jz(1-v^2)^{7/2}}{[vaz+(1-v^2)^{3/2}]^3}~.
\end{eqnarray}
Substituting (\ref{xdot}), (\ref{xprime}) and (\ref{seconderivatives}) into the equation of motion (\ref{eom}), it is easy to verify that it is in fact satisfied, thus providing an independent check of the extremization argument of \cite{mikhailov}.

\subsection{Radiation damping}\label{dampingsubsec}

When the quark/endpoint moves at constant velocity $v$, (\ref{mikhsol}) describes an upright (i.e., purely radial) string translating uniformly, as one could have anticipated by Lorentz invariance. For any accelerated motion, an event horizon  develops on the string worldsheet \cite{dragtime}, associated with the expected fact that the quark emits gluonic radiation and is thus subjected to a corresponding damping force. In \cite{lorentzdirac,damping} it was shown, for general three-dimensional motions, that this physics is correctly captured by the forced boundary condition (\ref{stringbc}). Indeed, through use of Mikhailov's embedding, this condition can be rewritten purely in terms of the time evolution of the endpoint, and so can be interpreted as an equation of motion for the quark, incorporating Lorentz-covariant formulas for its rate of radiation and intrinsic four-momentum. We will here reobtain this information via a streamlined procedure adapted to our one-dimensional setting.

For the radiation damping to be noticeable at the level of the quark equation of motion, we need the quark to have a finite mass, so from now on we allow again $z_m>0$ (which, as we
emphasized in Section 2, inevitably means that our non-Abelian source is no longer pointlike but has size
$z_m$).   The novelty is then that the boundary condition (\ref{quarkx}) is now imposed not at the AdS boundary but at the radial position $z=z_m$ where the string endpoint lies. As before, this condition by itself does not pick out a unique string embedding. Just like we discussed in the previous subsection for the infinitely massive case, we additionally require the solution to be `retarded' or `purely outgoing', in order to focus on the gluonic field causally set up by the quark. As in \cite{dragtime,lorentzdirac,damping}, we can inherit this structure by truncating a suitably selected retarded Mikhailov solution.
The  solutions of
interest to us can thus be regarded as the $z\ge z_m$ portions of the embeddings (\ref{mikhsol}),
which are parametrized by data at the AdS boundary $z=0$. {}From this point on we will use tildes to label
these (now merely auxiliary) data, and distinguish them from the actual physical quantities
(velocity, time, etc.) associated with the endpoint/quark at $z=z_m$, which will be denoted
without tildes.

In this notation, (\ref{mikhsol}) reads
\begin{eqnarray}\label{mikhsoltilde}
X(\tilde{t}_r,z)&=&\tilde{x}+\frac{\tilde{v}z}{\sqrt{1-\tilde{v}^2}}~,\\
t(\tilde{t}_r,z)&=&\tilde{t}_r+\frac{z}{\sqrt{1-\tilde{v}^2}}~,\nonumber
\end{eqnarray}
and expressions (\ref{xdot}), (\ref{xprime}) and (\ref{seconderivatives}) continue to hold if their right-hand sides are similarly decorated with tildes. Using this information in (\ref{momenta}) we can deduce that
\begin{equation}
\Pi_x^z=\frac{\tilde{a}}{z(1-\tilde{v}^2)^{3/2}}~,
\end{equation}
which when evaluated at $z=z_m$ implies, with the aid of (\ref{stringbc}), that
\begin{equation}\label{f}
\frac{\tilde{a}}{z_m(1-\tilde{v}^2)^{3/2}}= \frac{2\pi}{\sqrt{\lambda}}F\equiv\slash{F}~.
\end{equation}
By evaluating (\ref{xdot}) at the endpoint we can also infer that
\begin{equation}\label{v}
v=\frac{\tilde{a}z_m+\tilde{v}(1-\tilde{v}^2)^{3/2}}{\tilde{v}\tilde{a}z_m
+(1-\tilde{v}^2)^{3/2}}~,
\end{equation}
which we can invert in combination with (\ref{f}) to learn that
\begin{eqnarray}\label{tildes}
\tilde{v}&=&\frac{v-z_m^2\slash{F}}{1-z_m^2v\slash{F}}\\
\tilde{a}&=&z_m\slash{F}\frac{(1-v^2)^{3/2}(1-z_m^4\slash{F}^2)^{3/2}}{(1-z_m^2v\slash{F})^3}~.\nonumber
\end{eqnarray}

Having already incorporated the dynamical information contained in (\ref{stringbc}) and (\ref{mikhsol}), the equation controlling the evolution of the quark now follows simply by substituting (\ref{tildes}) into the simple kinematical statement
\begin{equation}\label{kinematiceom}
\frac{d}{d\tilde{t}_r}(\tilde{\gamma}\tilde{v})={\tilde{\gamma}}^3{\tilde{a}}~.
\end{equation}
Reading off from (\ref{mikhsol}) that
\begin{equation}\label{tr}
t_r\equiv t(\tilde{t}_r,z_m)=\tilde{t}_r+\frac{z_m}{\sqrt{1-\tilde{v}^2}}~,
\end{equation}
we can deduce with the aid of (\ref{tildes}) that
\begin{equation}
\frac{d t_r}{d\tilde{t}_r}=\frac{(1-\tilde{v}^2)^{3/2}+\tilde{v}\tilde{a}z_m}{(1-\tilde{v}^2)^{3/2}}
=\frac{1-z_m^4\slash{F}^2}{1-z_m^2v\slash{F}}~.
\end{equation}
Together with (\ref{tildes}), this turns (\ref{kinematiceom}) into
\begin{equation}
\frac{d}{d t_r}\left(\frac{\gamma v-z_m^2\gamma\slash{F}}{\sqrt{1-z_m^4\slash{F}^2}}\right)
=z_m\slash{F}\frac{(1-z_m^2v\slash{F})}{(1-z_m^4\slash{F}^2)}~,
\end{equation}
or, equivalently,
\begin{equation}\label{peom}
{d\over dt}\left(\frac{m\gamma v-{\sqrt{\lambda}\over 2\pi m}\gamma F }{\sqrt{1-{\lambda\over
 4\pi^2 m^4}F^2}}\right)=F -{\sqrt{\lambda}\over 2\pi} {F^2\over m^2}\left(\frac{v-{\sqrt{\lambda}\over 2\pi m^2}F
 }{1-{\lambda\over 4\pi^2 m^4}F^2}\right)~,
 \end{equation}
which is the quark equation of motion we were after. As explained in \cite{dragtime,lorentzdirac,damping}, it has a transparent and pleasant interpretation in gauge theory language: recognizing
\begin{equation}\label{pq}
p_q\equiv\frac{m\gamma v-\frac{\sqrt{\lambda}}{2\pi m}\gamma F}{\sqrt{1-\frac{\lambda}{4\pi^2m^4}F^2}}
\end{equation}
as the intrinsic momentum of the quark and
\begin{equation}\label{prad}
{d P_{\mbox{\scriptsize rad}}\over
dt}={\sqrt{\lambda}\over 2\pi} {F^2\over m^2}\left(\frac{v-{\sqrt{\lambda}\over 2\pi m^2}F
 }{1-{\lambda\over 4\pi^2 m^4}F^2}\right)
\end{equation}
as the rate at which momentum is carried away from it by the emitted gluonic radiation (and consequently, as the negative of the radiative damping force), (\ref{peom}) is seen to amount merely to the statement of momentum conservation,
\begin{equation}\label{psplit}
 {d P\over dt}\equiv {d p_q\over dt}+{d P_{\mbox{\scriptsize rad}}\over
 dt}=F~.
 \end{equation}
The total momentum supplied to the system by the external force must either increase the intrinsic momentum of the quark or be radiated away into the gluonic field. The fact that this eminently physical gauge-theoretic statement has arisen so straightforwardly from the standard string dynamics is a nice illustration of the power of the AdS/CFT correspondence.

 Needless to say, the total energy is also conserved, and the corresponding statement can be obtained by applying the above argument to $d\tilde{\gamma}/{d\tilde{t}_r}=\tilde{\gamma}^3\tilde{v}\tilde{a}$ instead of (\ref{kinematiceom}), yielding
  \begin{equation}\label{eeom}
{d\over dt}\left(\frac{m\gamma -{\sqrt{\lambda}\over 2\pi m}\gamma v F }{\sqrt{1-{\lambda\over
 4\pi^2 m^4}F^2}}\right)
 =
 vF -{\sqrt{\lambda}\over 2\pi} {F^2\over m^2}\left(\frac{1-{\sqrt{\lambda}\over 2\pi m^2}vF
 }{1-{\lambda\over 4\pi^2 m^4}F^2}\right)~.
 \end{equation}
Identifying
\begin{equation}\label{eq}
E_q\equiv\frac{m\gamma -\frac{\sqrt{\lambda}}{2\pi m}\gamma vF}{\sqrt{1-\frac{\lambda}{4\pi^2m^4}F^2}}
\end{equation}
as the intrinsic energy of the quark and
\begin{equation}\label{erad}
{d E_{\mbox{\scriptsize rad}}\over
dt}={\sqrt{\lambda}\over 2\pi} {F^2\over m^2}\left(\frac{1-{\sqrt{\lambda}\over 2\pi m^2}vF
 }{1-{\lambda\over 4\pi^2 m^4}F^2}\right)
\end{equation}
as the rate at which it radiates energy,
(\ref{eeom}) embodies the  energy conservation statement
\begin{equation}\label{esplit}
 {d E\over dt}\equiv {d E_q\over dt}+{d E_{\mbox{\scriptsize rad}}\over
 dt}=vF~.
 \end{equation}

  The Lorentz-covariant generalization of equations (\ref{peom}) and (\ref{eeom}), for arbitrary three-dimensional motions of the quark, can be found in \cite{lorentzdirac,damping}. As emphasized there, the result can be understood to be an extension of the well-known Lorentz-Dirac equation of classical electrodynamics \cite{dirac} to the strongly-coupled non-Abelian setting. Unlike its Lorentz-Dirac counterpart, the AdS/CFT extension is found to be non-linear and to have no pathological solutions.

We note in passing that (\ref{tildes}) and (\ref{tr}) allow us to eliminate all auxiliary quantities and rewrite the string embedding (\ref{mikhsol}) purely in terms of physical quark/endpoint data,
\begin{eqnarray}\label{mikhsolnotildes}
X(t_r,z)&=&x+\frac{(v-z_m^2\slash{F})(z-z_m)}
{\sqrt{1-v^2}\sqrt{1-z_m^4\slash{F}^2}}~,\\
t(t_r,z)&=&t_r+\frac{(1-z_m^2v\slash{F})(z-z_m)}
{\sqrt{1-v^2}\sqrt{1-z_m^4\slash{F}^2}}~.\nonumber
\end{eqnarray}

 The non-trivial form of the quark dispersion relation embodied in (\ref{pq}) and (\ref{eq}) is a direct consequence of its extended, non-pointlike, nature, which we took care to emphasize in Section \ref{quarksec}. Indeed, the characteristic length scale appearing in the quark dressing seen in (\ref{pq}) is none other than $z_m$, which we know to play the role of Compton wavelength. The AdS/CFT correspondence is thus teaching us precisely how the gluonic cloud surrounding the quark is deformed upon the application of an external force.

 For $F=0$, (\ref{pq}) reduces to the usual dispersion relation $p_q=m\gamma v$ (and similarly $E_q=m\gamma$). The same is true in the infinite-mass/pointlike limit $z_m\to 0$. A consequence of this last statement that we would like to stress as an important lesson of the above procedure is that, in terms of the auxiliary/tilde variables, the intrinsic quark momentum and energy always take the usual form, $p_q=m\tilde{\gamma} \tilde{v}$ (as seen in (\ref{kinematiceom})) and $E_q=m\tilde{\gamma}$, and the non-trivial dressing present in (\ref{pq}) and (\ref{eq}) arises only upon using (\ref{tildes}) to rewrite this in terms of the physical/nontilde variables. In the next section we will learn that this continues to hold true even in the finite-temperature setting.

 For later use it is also useful to recall that in \cite{mikhailov,dragtime}, a time-integrated version of the  momentum/energy split seen in (\ref{psplit}) and (\ref{esplit}) was shown to be achieved on the string theory side directly at the level of the total momentum/energy carried by the string worldsheet. In more detail, \cite{mikhailov} obtained the infinite-mass limit of (\ref{prad}) and (\ref{erad}) (which, surprisingly, has the same functional form as the Lienard formula of classical electrodynamics) by using (\ref{mikhsol}) in the integral of the  momentum/energy densities $\Pi^t_x$ and $\Pi^t_t$ given in (\ref{momenta}). These calculations disregarded a total derivative term, which was later shown in \cite{dragtime} to correspond to the infinite-mass limit of (\ref{pq}), (\ref{eq}). The same reference extended the analysis to the case of finite mass. Of this story, the element that we most want to emphasize is that, at the worldsheet level, the intrinsic quark momentum and energy (\ref{pq}) and (\ref{eq}) are obtained as surface terms, which are naturally evaluated at the string endpoint.

Our one-dimensional equation of motion (\ref{peom}) or (\ref{eeom}) can be rewritten in the form
 \begin{equation} \label{alinear}
a=\frac{z_m\slash{F}(1-v^2)^{3/2}}{\sqrt{1-z_m^4{\slash{F}}^2}}
+\frac{z^2_m\dot{\slash{F}}(1-v^2)}{1-z_m^4{\slash{F}}^2}~,
\end{equation}
from which it is clear that the quark will accelerate (and thereby radiate) only if externally forced. This is precisely as one would expect for a particle in vacuum, but of course the situation will be different when we introduce a thermal medium in the next section.

\section{Finite Temperature}\label{finitetempsec}

We will now extend the results of the previous section to the case where the quark moves inside a thermal MSYM plasma. In Section \ref{embeddingtempsubsec} we will obtain the first finite-temperature correction to the string embedding, and then in Section \ref{dampingtempsubsec} we will use this information to infer the rate at which the quark is damped by the plasma.

\subsection{First thermal correction to string embedding}\label{embeddingtempsubsec}
When we turn the temperature back on we have $z_h=1/\pi T<\infty$, and the momenta (\ref{momenta})
imply that the string equation of motion (\ref{ngeom}) now takes the form
\begin{equation}
\frac{\partial}{\partial t}\left(\frac{\dot{X}}{z^2 h\sqrt{1+h X'^2-\dot{X}^2/h}}\right)-\frac{\partial}{\partial z}\left(\frac{h X'}{z^2\sqrt{1+h X'^2-\dot{X}^2/h}}\right)=0~,
\end{equation}
or, equivalently,
\begin{equation}\label{ngeomtemp}
\begin{split}
0=&\;\ddot{X}-h^2X''+h\ddot{X}X'^2+hX''\dot{X}^2-2h\dot{X}X'\dot{X}'+\frac{2h^2X'}{z}+\frac{2h^3X'^3}{z}-\frac{2hX'\dot{X}^2}{z}\\
&\;-hh'X'-\begin{matrix} \frac{1}{2}
\end{matrix}h^2h'X'^3+\begin{matrix} \frac{3}{2}
\end{matrix}h'X'\dot{X}^2~.
\end{split}
\end{equation}

It would be highly desirable to obtain from this finite-temperature equation a family of exact solutions analogous to (\ref{mikhsol}), but unfortunately, we have not been able to accomplish this. Our aim will be instead to derive the first corrections to the string/quark dynamics induced by the presence of the thermal medium. As explained in the Introduction, when a heavy quark ($m\gg\sqrt{\lambda}T/2$) is created within the MSYM plasma we expect its initial evolution to be quite close to that in vacuum, with the thermal effects only starting to be felt gradually, as the gluonic field set up by the quark begins to interact with the medium and thereby induce a damping effect on the quark. In the string theory description, this corresponds to the fact that (for $z_m/z_h\ll 1$) a wave propagating upward from the string endpoint is initially far from the Schwarzschild-AdS horizon, and so evolves essentially as in pure AdS, with small corrections arising gradually as the wave moves
 deeper into the bulk. As long as we restrict attention to the early-time evolution where the disturbed portion of the string  remains far from the black hole horizon,
\begin{equation}\label{smalltemp}
\left(z\over z_h\right)^4=\pi^4 T^4 z^4 \ll 1~,
\end{equation}
the form of the metric (\ref{metric}) enables us to regard the string profile as a small perturbation about the zero-temperature solution (\ref{mikhsol}). By the UV/IR connection \cite{uvir}, $z$ corresponds to a length scale $d=z$
on the MSYM side, so the condition (\ref{smalltemp}) states that the region of the gluonic field that is disturbed by the motion of the quark remains small in units of the inverse temperature of the plasma.
With this restriction in mind, we thus expand the exact equation of motion (\ref{ngeom}) in powers of the small parameter (\ref{smalltemp}) and keep only the leading non-trivial correction:
\begin{equation}
\begin{split}\label{smalltempngeom}
0=&\;\left(\ddot{X}-X''+\ddot{X}X'^2+X''\dot{X}^2-2\dot{X}X'\dot{X}'+\frac{2X'}{z}+\frac{2X'^3}{z}-\frac{2X'\dot{X}^2}{z}\right)\\
&\;+\pi^4T^4z^4\left(2X''-\ddot{X}X'^2-X''\dot{X}^2+2\dot{X}X'\dot{X}'
-\frac{4X'\dot{X}^2}{z}-\frac{4X'^3}{z}\right)+\mathcal{O}(\pi^8T^8z^8)~.
\end{split}
\end{equation}
In parallel with this, we write the string embedding as a perturbation of Mikhailov's vacuum profile (\ref{mikhsol}), which we hereby denote with a ``$0$'' subindex,
\begin{equation}
X(t,z)=X_0(t,z)+Y(t,z)~,
\end{equation}
and where it is understood that $|Y/X_0|\ll 1$ throughout the region of interest. Plugging this ansatz into (\ref{smalltempngeom}), and keeping only terms up to first order in the perturbation $Y$, we obtain the linearized equation of motion
\begin{equation}\label{linearngeom}
A\dot{Y}+BY'+C\ddot{Y}+DY''+E\dot{Y}'+F=0
\end{equation}
where
\begin{eqnarray}\label{linearngeomcoeffs}
A&=&2X''_0\dot{X}_0-2X'_0\dot{X}'_0-\frac{4X'_0\dot{X}_0}{z}~,\nonumber\\
B&=&2\ddot{X}_0X'_0-2\dot{X}_0\dot{X}'_0+\frac{2}{z}+\frac{6{X'_0}^2}{z}-\frac{2{\dot{X}_0}^2}{z}~,\nonumber\\
C&=&{X'_0}^2+1~,\nonumber\\
D&=&{\dot{X}_0}^2-1~,\\
E&=&-2\dot{X}_0X'_0~,\nonumber\\
F&=&\pi^4T^4z^4\left(2X''_0-\ddot{X}_0{X'_0}^2
-X''_0{\dot{X}_0}^2+2\dot{X}_0X'_0\dot{X}'_0
-\frac{4X'_0{\dot{X}_0}^2}{z}-\frac{4{X'_0}^3}{z}\right)~.\nonumber
\end{eqnarray}

The coefficients  obtained after substituting (\ref{xdot}), (\ref{xprime}) and (\ref{seconderivatives}) into (\ref{linearngeomcoeffs}) have a very complicated dependence on $\tilde{v}$, $\tilde{a}$, $\tilde{j}$ and $z$, so it is challenging to find a full solution of even the linearized equation (\ref{linearngeom}). We therefore need to make some sort of additional simplification. Guided again by phenomenological considerations, we choose to restrict attention to a quark that is ultra-relativistic. More specifically, we assume that
\begin{equation}\label{ultrarel}
\alpha\equiv\sqrt{1-\tilde{v}^2}\ll 1~,
\end{equation}
which we will be able below to reexpress directly in terms of the physical quark velocity $v$. We can then proceed to expand (\ref{linearngeomcoeffs}) in a power series in this additional small parameter, using
$\tilde{v}=\sqrt{1-\alpha^2}\simeq 1-\begin{matrix}\frac{1}{2}\end{matrix}\alpha^2$, $\tilde{v}^2=1-\alpha^2$, $\tilde{v}^3\simeq1-\begin{matrix}\frac{3}{2}\end{matrix}\alpha^2$, etc. As we will find shortly, the expansion of the embedding will actually end up proceeding in powers of $\alpha^2$. Notice that,
 to the extent that we wish to consider successive corrections in $\alpha^2$ while continuing to disregard the higher-order thermal corrections, we are restricted to the regime where
\begin{equation}\label{smallparameters}
\pi^4 T^4 z^4 \ll 1-\tilde{v}^2\ll 1~,
\end{equation}
which as we will see in the next section is in fact consistent with the phenomenological setup at RHIC.

Now, in performing the ultrarelativistic expansion of (\ref{linearngeomcoeffs}), given the form of the denominators in (\ref{xdot}), (\ref{xprime}), (\ref{seconderivatives}), it becomes necessary to establish whether the quantity $\tilde{a}z$, which we expect to be small for a thermally-induced acceleration, is smaller than, of the same order as, or larger than the small quantity $(1-\tilde{v}^2)^{3/2}$. We have explored all three possibilities and found that the only one that is self-consistent is
\begin{equation}\label{con2}
\frac{\tilde{a}z}{(1-\tilde{v}^2)^{3/2}}\ll 1~.
\end{equation}
Notice that this constraint is additionally consistent with the facts that (i) we have already placed an \emph{upper} bound on $z$ through (\ref{smalltemp}); (ii) we need the acceleration not to be to large in order to remain within the ultrarelativistic regime (\ref{ultrarel}); (iii) for reasons to be explained in Section \ref{phenosec}, we will need the disturbance on the string to stay below the finite-temperature generalization of the dynamical worldsheet horizon that at zero-temperature lies close to a stationary limit curve \cite{dragtime} located at $z_{\mbox{\scriptsize ergo}}\simeq {(1-\tilde{v}^2)^{3/2}}/\tilde{a}$. An additional observation similar in spirit to this last remark is that, in the constant-velocity thermal embedding of \cite{hkkky,gubser}, $z_v\equiv (1-\tilde{v}^2)^{1/4}z_h$ marks the location of a stationary limit curve that is simultaneously a worldsheet horizon \cite{gubserqhat,ctqhat}, and our condition (\ref{smallparameters}) restricts us to radial locations below this.

Combining these assumptions we find that the coefficients (\ref{linearngeomcoeffs}) reduce to
\begin{eqnarray}\label{linearsimpcoeffs}
A&=&1+\alpha^2~,\nonumber\\
B&=&\alpha^2~,\nonumber\\
C&=&\frac{2\alpha}{z}~,\\
D&=&\frac{2\alpha^2}{z}~,\nonumber\\
E&=&2\alpha~,\nonumber\\
F&=&3\pi^4T^4z^3\alpha~.\nonumber
\end{eqnarray}
It is natural then to propose a series expansion for the perturbation on the string embedding,
\begin{equation}\label{yexpansion}
Y(t_r,z)=\sum_k f_k(z)\alpha^k~,
\end{equation}
where the functions $f_k(z)$ could generically also depend on $t_r$ through $\tilde{a}$, $\tilde{j}$, etc.  Upon taking the corresponding derivatives in (\ref{smalltempngeom}) and matching coefficients of a given order in $\alpha$, it is easy to see that the terms with $k<-1$ lead to homogeneous differential equations for $f_k$, meaning that these terms do not contain any information about the thermal medium, and can thus be set to zero. (They must in fact be set to zero if we are to preserve the purely-outgoing structure of the string profile.) The leading order term in (\ref{yexpansion}) is then of order $1/\alpha$, which as seen in (\ref{mikhsoltilde}) is the same velocity dependence present in the unperturbed  solution $X_0$.

The $k=-1$ coefficient is found to satisfy the differential equation
\begin{equation}
2f_{-1}'(z)+\pi^4T^4z^4=0~,
\end{equation}
whose general solution is
\begin{equation}
f_{-1}(z)=-\frac{\pi^4T^4z^5}{10}+c~.
\end{equation}
Demanding that $Y(t,z=0)=0$, so that $\tilde{x}(\tilde{t})$ continues to represent the trajectory of the (auxiliary or physical, depending on whether or not $z_m=0$) string endpoint, we are forced to set $c=0$, and conclude that, to this order in our small-temperature and large-velocity expansion,
\begin{equation}\label{ylowest}
Y(\tilde{t}_r,z)=-\frac{\pi^4T^4z^5}{10\sqrt{1-\tilde{v}^2}}~.
\end{equation}
Notice that $|Y/X_0|$ is consistently small, of order (\ref{smalltemp}), as expected from the size of the inhomogeneous term in the equation (\ref{linearngeom}).

Starting from (\ref{ylowest}), we can recursively obtain higher-order corrections to the perturbation (\ref{yexpansion}). The $k=0$ coefficient is found to obey
\begin{equation}
f_{0}'(z)=0~,
\end{equation}
implying that
\begin{equation}
f_{0}(z)=0~.
\end{equation}
At level $k=1$ we find
\begin{equation}
2f_{1}'(z)+\frac{3}{2}\pi^4T^4z^4=0~,
\end{equation}
whose solution is
\begin{equation}
f_{1}(z)=-\frac{3\pi^4T^4z^5}{20}~.
\end{equation}
The next coefficient, $f_2$, is again found to vanish, so, up to terms of order $\alpha^3$, we learn that
\begin{equation}
Y(\tilde{t}_r,z)=-\frac{\pi^4T^4z^5}{10\sqrt{1-\tilde{v}^2}}
-\frac{3\pi^4T^4z^5\sqrt{1-\tilde{v}^2}}{20}~,
\end{equation}
and therefore
\begin{equation}\label{xcorrected2}
X(\tilde{t}_r,z)=\tilde{x}+\frac{\tilde{v}z}{\sqrt{1-\tilde{v}^2}}
-\frac{\pi^4T^4z^5}{10\sqrt{1-\tilde{v}^2}}-\frac{3\pi^4T^4z^5\sqrt{1-\tilde{v}^2}}{20}~.
\end{equation}
 One can continue this expansion to higher order in $\alpha$, and upon doing so one finds that terms involving $\tilde{a}$ and $\tilde{j}$ begin to appear in the equation of motion, indicating the need for a more general ansatz. We will not pursue this further here, however, as our main interest lies in determining the leading thermal damping, which as we will now show is available to us already from the first $T$-dependent term in (\ref{xcorrected2}).

\subsection{Early-time thermal damping}\label{dampingtempsubsec}

We now use the thermally-corrected string embedding derived in the previous subsection to learn about the way in which the MSYM plasma begins to damp the quark. Including only the leading nontrivial correction in the small quantities (\ref{smalltemp}) and (\ref{ultrarel}), we have learned that
\begin{equation}\label{xcorrected1}
X(\tilde{t}_r,z)=\tilde{x}+\frac{\tilde{v}z}{\sqrt{1-\tilde{v}^2}}
-\frac{\pi^4T^4z^5}{10\sqrt{1-\tilde{v}^2}}~.
\end{equation}
{}From this it follows that
\begin{equation}\label{xdottemp}
\dot{X}(\tilde{t}_r,z)\equiv\left(\p X\over\p t\right)_z=
\frac{\tilde{v}(1-\tilde{v}^2)^{3/2}+\tilde{a}z
\left(1-\frac{1}{10}\pi^4T^4 z^4\tilde{v}\right)}
{\tilde{v}\tilde{a}z+(1-\tilde{v}^2)^{3/2}}
\end{equation}
and
\begin{equation}\label{xprimetemp}
X^{\prime}(\tilde{t}_r,z)\equiv\left(\p X\over\p z\right)_t=
\frac{-\frac{1}{2}\pi^4T^4z^4(1-\tilde{v}^2)^{3/2}-\tilde{a}z
\left(1-\tilde{v}^2+\frac{2}{5}\pi^4T^4 z^4\tilde{v}\right)}
{\tilde{v}\tilde{a}z\sqrt{1-\tilde{v}^2}+(1-\tilde{v}^2)^{2}}~,
\end{equation}
which evaluated at the string endpoint yield expressions for the physical quark/endpoint velocity
$v=\dot{X}(\tilde{t}_r,z_m)$ and the endpoint slope $s\equiv X^{\prime}(\tilde{t}_r,z_m)$.

Of particular phenomenological relevance is the case where the quark is not externally forced, where we can see
from (\ref{stringbc}) and (\ref{momenta}) that the string satisfies the familiar free-end (Neumann) boundary condition $s=0$.
We can then invert (\ref{xdottemp}) and (\ref{xprimetemp}) evaluated at $z=z_m$ to find that
\begin{equation}\label{vtildetemp}
\tilde{v}=v+\begin{matrix}\frac{1}{2}\end{matrix}\pi^4T^4z_m^4
\end{equation}
and
\begin{equation}\label{atildetemp}
\tilde{a}=-\frac{5\pi^4T^4z_m^3
\left[1-\left(v+\frac{1}{2}\pi^4T^4z_m^4\right)^2\right]^{3/2}}
{10(1-v^2)-6\pi^4T^4z_m^4v-\frac{1}{2}\pi^8T^8z_m^8}~.
\end{equation}

It is now useful to remember that, for arbitrary $z_m$, our calculational framework interprets the physical string as merely the ($z\ge z_m$) portion of an auxiliary string that reaches all the way down to the Schwarzschild-AdS boundary. In this interpretation,  $\tilde{v}$ denotes the velocity of the (fictitious) endpoint at $z=0$, while $v$ represents the velocity of that same embedding at radial depth $z=z_m$. This implies that we can move along the physical string by fixing a value of $\tilde{v}$ (to select a particular embedding) and considering $z_m$ as the free parameter $z$ that is greater than or equal to the value (\ref{zm}) determined by the quark mass. We see in (\ref{vtildetemp}) that, in this reading, the velocity $v$ decreases as $z_m$ is increased, which tells us that, due to the presence of the black hole, the horizontal velocity of the string is smaller for points deeper into the Schwarzschild-AdS bulk. In other words, whereas at zero temperature an externally unforced string would be upright, we find here that the thermal effects make the string lean back, and therefore exert a damping force on its endpoint, just as one would expect, and as is visible (for the auxiliary endpoint) also in (\ref{atildetemp}).

Beyond the anticipated fact that the string leans back, the precise form of the relation we have obtained in (\ref{vtildetemp}) is in fact directly associated with an important physical feature of motion in a strongly-coupled plasma. To see this, note first that the requirement that the Nambu-Goto action (\ref{nambugoto}) describing a string embedded in the black hole metric (\ref{metric}) be real (i.e., that the string
worldsheet be timelike) imposes a bound on classically realizable embeddings, which generalizes the familiar condition that a particle not be able to exceed the speed of light. The precise general bound takes the form \cite{dragtime}
\begin{equation}\label{ngbound1}
\left({\p\vec{X}\over \p t}\right)^2\le h\frac{1+h\left({\p\vec{X}\over \p z}\right)^2}
{1+h\left({\p\vec{X}\over \p z}\right)^2\sin\vartheta }~,
\end{equation}
where $\vartheta\equiv\angle(\p \vec{X}/\p t,\p \vec{X}/\p z)$. For a string that moves and stretches
along a single direction $x$, as we are considering in the present paper, $\vartheta=0$ or $\pi$ and (\ref{ngbound1}) reduces to
\begin{equation}\label{ngbound2}
\left({\p X\over \p t}\right)^2\le h\left(1+h\left({\p X\over \p z}\right)^2\right)~.
\end{equation}

The constraint (\ref{ngbound2}), which is of course evident directly from (\ref{nambugotosimpl}), must be satisfied by the embedding function $X(t,z)$ at all instants and all points along the string. In particular, at the physical string endpoint we must have
\begin{equation}\label{ngbound3}
v^2\le h_m\left(1+h_m s^2\right)~,
\end{equation}
where $h_m\equiv h(z_m)$. In terms of the external force \cite{dragtime}, this reads
\begin{equation}\label{vmpi}
v\le \frac{h_m}{\sqrt{(h_m-z_m^4\slash{F}^2)(1+z_m^4\slash{F}^2)}} =\frac{v_m^2}{\sqrt{(v_m^2-\lambda
F^2/4\pi^2 m^4 []^4)(1+\lambda F^2/4\pi^2
m^4 []^4)}}~,
\end{equation}
where
\begin{equation}\label{vm}
v_m\equiv\sqrt{h_m}=\sqrt{1-\pi^4 T^4 z_m^4}~,
\end{equation}
and $[]$ denotes the expression within brackets in (\ref{zm}).

In the unforced case we are presently considering, this implies the simple velocity bound
\begin{equation}\label{vbound}
v\le v_m~.
\end{equation}
The appearance of (\ref{vm}) as a limiting velocity for quarks, mesons and other objects moving through a strongly-coupled thermal plasma is implicit
   in at least \cite{argyres1,gubserqhat,ctqhat}, and was discussed
   more explicitly from various perspectives in \cite{mateosvm,liu4,argyres3,dragtime,ms,st}.
The origin of the bound is easy to understand on the string theory side of the
duality: a coordinate velocity $v=v_m$ corresponds to a
\emph{proper} velocity $V$ equal to that of light at the position of the string endpoint,
$z=z_m$ \cite{argyres1}. The interesting feature is that it is $v$, and not $V$, that
corresponds to the gauge theory velocity. This yields then a nontrivial prediction of the AdS/CFT correspondence for strongly-coupled plasmas, which in turn might have interesting phenomenological consequences, such as the photon peak predicted in \cite{cm} or the Cherenkov emission of mesons analyzed in \cite{cdm}.

Now, for a heavy quark (consistent with (\ref{smalltemp})), the limiting velocity (\ref{vm}) reads
\begin{equation}\label{vmsimpl}
v_m=1-\frac{1}{2}\pi^4 T^4 z_m^4~,
\end{equation}
so we can finally recognize the physical significance of (\ref{vtildetemp}): it states simply that
\begin{equation}\label{vvtildesimpl}
 1-\tilde{v}=v_m-v\simeq 1-\frac{v}{v_m}~,
 \end{equation}
which shows that the bound (\ref{vbound}) is equivalent to the standard relativistic constraint  $\tilde{v}\le 1$.

The preceding discussion motivates rewriting (\ref{vtildetemp}) and (\ref{atildetemp}) in the  form
\begin{equation}\label{vtildetemp2}
\tilde{v}=\frac{v}{v_m}
\end{equation}
(a relation that is likely to be exact) and
\begin{equation}\label{atildetemp2}
\tilde{a}=-\frac{1}{2}\pi^4T^4z_m^3\sqrt{1-\left(\frac{v}{v_m}\right)^2}~,
\end{equation}
which are more physically transparent and, to the order of our analysis, equivalent.
Notice that the fact that
\begin{equation}\label{vtildev}
\sqrt{1-\tilde{v}^2}=\sqrt{1-\left(\frac{v}{v_m}\right)^2}~,
\end{equation}
implies that the predicted limiting velocity will be automatically incorporated into the kinematics of our finitely-massive quark. We will return to this point below.
Notice also that (\ref{atildetemp2}) and (\ref{vtildetemp}) lead to
\begin{equation}\label{con2validated}
\frac{\tilde{a}z_m}{(1-\tilde{v}^2)^{3/2}}
=\frac{\pi^4 T^4 z_m^4}{2(1-\tilde{v}^2)}~,
\end{equation}
which through (\ref{smallparameters}) shows that our assumption (\ref{con2}) is self-consistent, as promised.

We comment in passing that (\ref{tr}) and (\ref{vtildetemp2}) allow the embedding (\ref{xcorrected1}) to be rewritten in the form
\begin{eqnarray}\label{xcorrected1notildes}
X(t_r,z)&=&
x+\frac{v(z-z_m)}{\sqrt{v_m^2-v^2}}
-\frac{\pi^4T^4(z^5-z_m^5)v_m}{10\sqrt{v_m^2-v^2}}~,\nonumber\\
t(t_r,z)&=&t_r+\frac{(z-z_m)v_m}{\sqrt{v_m^2-v^2}}~,
\end{eqnarray}
where all auxiliary variables have been eliminated and only the physical quark data remain.
To avoid possible confusion, we should note that since the retarded time $t_r$ (where it is understood that $x$ and $v$ are to be evaluated) is defined just like in the unperturbed solution (\ref{mikhsolnotildes}), constant-$t_r$ curves on the worldsheet   are no longer null when finite-temperature contributions are taken into account. Given the  way in which Mikhailov's vacuum solution (\ref{mikhsol}) (or its finite mass generalization (\ref{mikhsolnotildes})) incorporates information flow on the worldsheet at the speed of light, it would be natural to define a thermally-corrected retarded time
 by following the shifted null curves. This can be done, and amounts to a reparametrization of the worldsheet, but we will refrain from showing the result because it is not particularly illuminating at our level of approximation, even though it should presumably be crucial for the structure of the full finite-temperature solution.

We now finally proceed to extract the information on the evolution of the physical endpoint/quark. {}From (\ref{tr}) we know that
\begin{equation}\label{dtr}
\left[1+\frac{\tilde{v}\tilde{a}z_m}{(1-\tilde{v}^2)^{3/2}}\right]d\tilde{t}_r=d t_r~.
\end{equation}
Combining this with (\ref{con2}), (\ref{vtildetemp2}) and the fact that $v_m$ is a constant,
we see that, to lowest nontrivial order in our double expansion in small quantities,
$$a\equiv \frac{dv}{dt_r}=v_m \frac{d\tilde{v}}{d\tilde{t}_r}=v_m\tilde{a}~.$$
Using (\ref{atildetemp2}), we deduce then that the quark/endpoint acceleration is given by
\begin{equation}\label{atemp}
a=-\frac{1}{2}\pi^4T^4z_m^3\sqrt{v_m^2-v^2}=-\frac{\pi\lambda^{3/2}T^4\sqrt{v_m^2-v^2}}{16m^3}~.
\end{equation}
This is the finite-temperature (and unforced) analog of (\ref{alinear}), and constitutes our main result. It shows that, as expected, the effect of the plasma is to slow down the quark, at an early-time rate that is naturally proportional to the $h-1$ correction present in the Schwarzschild-AdS metric (\ref{metric}), and therefore differs markedly from that implied by the stationary/late-time result
(\ref{elossgubser})-(\ref{plossgubser}).
We will elaborate on this difference in the next section.

The information contained in (\ref{atemp}) can be alternatively cast in terms of a rate of momentum or energy loss for the quark, by imitating the procedure employed in Section \ref{dampingsubsec}: using (\ref{vtildetemp2})-(\ref{dtr}), the kinematic identity (\ref{kinematiceom}),
$$
\frac{d}{d\tilde{t}_r}(\tilde{\gamma}\tilde{v})={\tilde{\gamma}}^3{\tilde{a}}~,
$$
 can be rewritten in the form
 \begin{equation}\label{peomtemp}
 \frac{d}{dt_r}\left(\frac{v}{\sqrt{v_m^2-v^2}}\right)=-\frac{\pi^4T^4z_m^3v_m^2}{2(v_m^2-v^2)}~,
 \end{equation}
 which is obviously equivalent to (\ref{atemp}). To interpret this in terms of momentum conservation, recall from Section \ref{dampingsubsec} that, at zero temperature, the intrinsic quark momentum always takes the standard Lorentz-invariant form in terms of the auxiliary/tilde variables. Since these variables are defined at the AdS boundary ($z=0$), where the presence of the black hole horizon is irrelevant, this same form must be preserved even in the finite-temperature context. In other words,  the thermal corrections to the quark dispersion relation arise not from a modification to the standard form $p_q\propto\tilde{\gamma}\tilde{v}$, but from the $T$-dependence present in the relations (\ref{vtildetemp2})-(\ref{atildetemp2}) that connect the tilde and non-tilde variables. We thus identify
 \begin{equation}\label{pqtemp}
 p_q\equiv m\tilde{\gamma}\tilde{v}=\frac{mv}{\sqrt{v_m^2-v^2}}
 \end{equation}
 as the intrinsic quark momentum, and recognize (\ref{peomtemp}) as the statement (\ref{psplit}) of momentum conservation, with the external force $F=0$ and the rate of momentum loss given by
 \begin{equation}\label{pradtemp}
 {d P_{\mbox{\scriptsize rad}}\over dt}=\frac{\pi^4T^4z_m^3 m v_m^2}{2(v_m^2-v^2)}
 =\frac{\pi\lambda^{3/2}T^4 v_m^2}{16m^2(v_m^2-v^2)}~.
 \end{equation}

 Actually, in the analysis of the previous paragraph we have ignored one subtlety. In Section \ref{dampingsubsec}
 we recalled from \cite{dragtime} the fact that the intrinsic quark momentum/energy arises from a worldsheet surface term. The same reference showed that, in the finite-temperature setting, this generally leads to an additional contribution from the \emph{upper} string endpoint, i.e., the one at the spacetime horizon $z=z_h$. This is a thermal correction that  was  argued in \cite{dragtime} to encode the initial/boundary conditions of the quark $+$ plasma system. For instance, in the case of a permanently static quark, the momentum surface term at the horizon vanishes, but the corresponding energy term gives the known $-\sqrt{\lambda}T/2$ correction to the quark rest mass. Precisely this same contribution appears in any quark configuration where the quark is assumed to have been at rest in the remote past, but the situation is different for other initial conditions (e.g., the case studied in \cite{hkkky,gubser} where the quark has constant velocity for all times \cite{dragtime}). For concreteness, we will here assume the resting initial condition, which is in any case the most natural way to make contact between our setup and the phenomenological context, as we will discuss in the next section. Since $-\sqrt{\lambda}T/2$ is independent of time, at the level of the time-differentiated intrinsic quark momentum/energy appearing in the conservation laws (\ref{psplit}) and (\ref{esplit}), only the contribution of the \emph{lower} ($z=z_m$) string endpoint matters, just as  we implicitly assumed in the preceding paragraph. Notice that, with our assumed initial conditions, the undifferentiated expression (\ref{pqtemp}) is correct as it stands.

Bearing this point in mind, just like in Section \ref{dampingsubsec} we can reinterpret the kinematic identity $d\tilde{\gamma}/d\tilde{t}_r=\tilde{\gamma}^3\tilde{v}\tilde{a}$ as the energy conservation law (\ref{esplit}), with $F=0$,
\begin{equation}\label{eqtemp}
 E_q\equiv m\tilde{\gamma}-\frac{\sqrt{\lambda}}{2\pi z_h}=\frac{mv_m}{\sqrt{v_m^2-v^2}}-\frac{1}{2}\sqrt{\lambda}T
 \end{equation}
 the intrinsic quark energy, and
 \begin{equation}\label{eradtemp}
 {d E_{\mbox{\scriptsize rad}}\over dt}=\frac{\pi^4T^4z_m^3m v_m v}{2(v_m^2-v^2)}
 =\frac{\pi\lambda^{3/2}T^4 v_m v}{16m^2(v_m^2-v^2)}
 \end{equation}
as the rate of energy loss for the quark. Equations (\ref{pradtemp}) and (\ref{eradtemp}) are the results that we advertised in the Introduction.

As noted below (\ref{vtildev}), we expect the quark kinematics (in the unforced case) to incorporate  (\ref{vm}) as a limiting velocity, and this is indeed what is seen in (\ref{pqtemp}) and (\ref{eqtemp}): the intrinsic momentum/energy of the quark diverges as $v\to v_m$.\footnote{Of course, within our approximation scheme (\ref{smalltemp}), we cannot get too close to this divergence. Our derivation does seem to suggest, however, that (\ref{vtildetemp2}), and therefore the divergence, is exact.}
Notice also that, due to the connection with the auxiliary/tilde variables, for any type of initial condition, the thermally-corrected dispersion relation for the quark always inherits the Lorentz structure
\begin{equation}\label{drtemp}
(E_q-E_h)^2=(p_q-p_h)^2+m^2~,
\end{equation}
where $E_h$ and $p_h$ denote the corresponding horizon surface terms. This is in fact true even  in the presence of external forcing, where, as studied in \cite{dragtime},  the finite-temperature expressions for the quark intrinsic energy and momentum have to incorporate the $F$-dependence seen already in the vacuum expressions (\ref{eq}) and (\ref{pq}). This dependence is in particular responsible for a physical divergence at the critical value of the force associated with quark-antiquark pair creation \cite{dragtime,lorentzdirac,damping}.
Needless to say, it would be interesting to repeat the above derivation in the case with external force $F\neq 0$, to deduce a thermally-corrected equation of motion that should presumably incorporate the full $F$-dependent bound (\ref{vmpi}). Unfortunately, in the general case it is difficult to invert the expressions relating the tilde and untilde quantities.

The equation of motion we have derived for the quark moving through the plasma, be it in the form (\ref{atemp}), (\ref{psplit}) or (\ref{esplit}), admits an exact solution where $v(t)$ is a cosine, of which only the linear portion of the evolution,
\begin{equation}\label{vsoltemp}
v(t)=v_0-\frac{\pi\lambda^{3/2}T^4}{16m^3}\sqrt{v_m^2-v_0^2}\,t~,
\end{equation}
is within the regime consistent with our approximations. More precisely, to stay within our ultra-relativistic approximation (\ref{ultrarel}), we are limited to consider times such that the second term in (\ref{vsoltemp}) is much smaller than the first, i.e.,
\begin{equation}\label{tultrarel}
t\ll t_{\mbox{\scriptsize ultra}}\equiv \frac{2}{\pi^4T^4 z_m^3\sqrt{v_m^2-v_0^2}}
=\frac{16m^3}{\pi\lambda^{3/2}T^4\sqrt{v_m^2-v_0^2}}~.
\end{equation}
Notice also that, in view of (\ref{xcorrected1notildes}), to satisfy the small-disturbance condition (\ref{smalltemp}) we must also have
\begin{equation}\label{ttemp}
t\ll t_{\mbox{\scriptsize temp}}\equiv \frac{1}{\pi T \sqrt{v_m^2-v_0^2}}~,
\end{equation}
which is more restrictive than (\ref{tultrarel}), because (\ref{smalltemp}) itself implies that
$t_{\mbox{\scriptsize temp}}\ll t_{\mbox{\scriptsize ultra}}$.

\section{Phenomenological Estimates}\label{phenosec}
We will now try to make some quantitative inferences from our results in the previous section. For applications
to the phenomenology of heavy ion collisions, we must choose values of the mass
parameter $z_m$ based on the charm and bottom quark masses, $m\simeq 1.4,4.8$ GeV. The issue
of how best to translate between the SYM and QCD parameters has been discussed in
\cite{gubsercompare} (see also \cite{bhm}). Taking $\alpha_{QCD}\simeq 0.5$ ($g_{QCD}\simeq\sqrt{2\pi}$), $N_c=3$ and
$T_{QCD}\simeq 250$ MeV, and employing the ``obvious'' prescription $g_{YM}=g_{QCD}$ ($g^2_{YM}N_c\simeq 19$) and
$T_{SYM}=T_{QCD}$, we find from (\ref{zm}) that $z_m/z_h\simeq 0.40$ for charm and
$z_m/z_h\simeq 0.11$ for bottom. If, on the other hand, one uses the ``alternative'' scheme
$g_{YM}^2 N_c\simeq 5.5$ (motivated in \cite{gubsercompare} through a rough matching of the
AdS/CFT and lattice quark-antiquark potentials) and $T_{SYM}=3^{-1/4}T_{QCD}\simeq 190$ MeV
(which follows from equating the energy densities of the two theories), then (\ref{zm})
leads to $z_m/z_h\simeq 0.16$ for  charm and $z_m/z_h\simeq 0.046$ for bottom.
For definiteness, and knowing that the validity of our approximations would only improve if we used the bottom mass, we will take representative values in the neighborhood of the charm mass,
$z_m/z_h=\pi T z_m \simeq 0.2$-$0.4$, which is the same range studied numerically in \cite{dragtime}. We then have $\pi^4 T^4 z_m^4~\simeq 2\times 10^{-3}$ - $2\times 10^{-2}$,
which is surely in compliance with (\ref{smalltemp}).

Next, we consider the velocities of the relevant quarks. For the mass values assumed above, the limiting velocity (\ref{vm}) works out to $v_m\simeq 0.987$-$0.999$. At RHIC, the presence of a parent hadron containing a single charm or bottom quark (i.e., a $D$ or $B$ meson, respectively) is inferred from the single electron resulting from its semileptonic decay \cite{heavyquarkexp}. Such electrons have been measured with transverse momenta up to $p_T\simeq 9$-10 GeV, which, in the case of a charm decay electron, suggests that we may take $\gamma\simeq 5-7$ as representative values for the Lorentz factors. These correspond to quark velocities $v\simeq 0.980$-$0.990$, which are indeed not too far from saturating the bound (\ref{vbound}). Using (\ref{vtildetemp2}), we thus have $1-\tilde{v}^2=1-(v/v_m)^2\simeq$ (2-4)$\times 10^{-2}$, in accord with our ultrarelativistic condition (\ref{ultrarel}), and (at least in the ``alternative'' setup) respecting the ordering (\ref{smallparameters}) of the small parameters employed in the double expansion within which we have been able to derive  analytic results. The numbers for LHC work out differently, since one expects heavy quarks there with transverse momenta up to $\sim 100$ GeV, but in any case, as noted in the Introduction, for extremely energetic quarks it will probably not be appropriate to carry out the entire energy loss calculation within a (weakly-curved) gravity dual.

Having verified the pertinence of our approximation scheme, we are now set to explore the consequences of our results. The main question is how our early rate of momentum or energy loss (\ref{pradtemp}) or (\ref{eradtemp}) compares to the corresponding late/stationary rates (\ref{plossgubser}) or (\ref{elossgubser}), under conditions relevant to the quark gluon plasma as produced at RHIC. By taking the ratio between the two sets of expressions, we see that
\begin{eqnarray}\label{latevsearly}
\frac{\left({d E_{\mbox{\tiny rad}}\over dt}\right)_{\mbox{\scriptsize late}}}
{\left({d E_{\mbox{\tiny rad}}\over dt}\right)_{\mbox{\scriptsize early}}}
=v_m\frac{\left({d P_{\mbox{\tiny rad}}\over dt}\right)_{\mbox{\scriptsize late}}}
{\left({d P_{\mbox{\tiny rad}}\over dt}\right)_{\mbox{\scriptsize early}}}
&\simeq&\frac{8m^2 v}{\lambda T^2}\sqrt{1-v^2}
\simeq2\sqrt{\frac{1-v^2}{\pi^4 T^4 z_m^4}}
\simeq 2\,\mbox{-}\,6~.
\end{eqnarray}
We see here that the initial damping proceeds at a rate that is a few times \emph{smaller} than the asymptotic rate determined in \cite{hkkky,gubser,ct}. This is consistent with the results obtained in \cite{dragtime} by studying the string evolution numerically, even though the analysis there was confined to nonrelativistic quark velocities. Given that the quark evolution at early times would be expected to be insensitive to the spatial extent of the plasma, it is interesting to note that a recent study \cite{cpz} found that  finite-size effects (relevant, by definition, at late times) also reduce the rate of energy loss as compared to the late-time infinite-medium results \cite{hkkky,gubser,ct}. By way of contrast, the finite-mass correction to the stationary result of \cite{hkkky,gubser} deduced in \cite{beuf} turned out to increase the  rate of energy loss as compared to (\ref{elossgubser}).

The appearance of $T^4$ in (\ref{eradtemp}), (\ref{pradtemp}) and (\ref{atemp}), which is saliently different from the $T^2$-dependence seen in (\ref{elossgubser}) and (\ref{plossgubser}), and in our derivation clearly stems from the $h-1$ deviation from pure AdS in the metric (\ref{metric}), is in consonance with an estimate for early-time energy loss presented in \cite{dominguez}, based on a heuristic argument involving saturation physics. By dimensional analysis in our conformally invariant theory, this scaling with temperature inevitably implies the dependence on $z_m^3$ seen in (\ref{eradtemp}), (\ref{pradtemp}) and (\ref{atemp}), which in turn fixes not only the mass scaling but also the peculiar factor of $\lambda^{3/2}$.

Beyond the fact that the early-time damping rate obtained in this paper is smaller than its asymptotic/stationary counterpart, (\ref{latevsearly}) displays another interesting feature: the functional dependence on the quark velocity is substantially different in the two cases. Whereas in the late-time regime the friction coefficient defined through
\begin{equation}\label{mu}
\mu\equiv -\frac{1}{p_q}\frac{dp_q}{dt}=\frac{1}{p_q}{d P_{\mbox{\scriptsize rad}}\over dt}
\end{equation}
is momentum-independent \cite{hkkky,gubser},
\begin{equation}\label{mulate}
\mu_{\mbox{\scriptsize late}}=\frac{\pi\sqrt{\lambda}T^2}{2m}~,
\end{equation}
in our early-time setup we find instead from (\ref{pqtemp}) and (\ref{pradtemp}) that
\begin{equation}\label{muearly}
\mu_{\mbox{\scriptsize early}}=\frac{\pi\lambda^{3/2}T^4v_m^2}
{16m^3v\sqrt{v_m^2-v^2}}
=\frac{\pi\lambda^{3/2}T^4(p_q^2+m^2)}
{16m^4 p_q}~,
\end{equation}
which given our restriction to the ultrarelativistic regime states that the friction coefficient grows linearly with the quark momentum. This growth might seem counterintuitive, but notice that (\ref{atemp}) expresses the expected physical property that a faster quark is \emph{less} damped by the plasma.\footnote{A growing friction coefficient was also found in \cite{liustirring}, for different physical reasons.}
 When translated to the nuclear modification factor $R_{AA}(p_T)$ (and similar observables) as in \cite{horowitz,hg,mr,kharzeev}, (\ref{muearly}) would imply a $p_T$-dependence that differs from that of both perturbative QCD and the late-time trailing string result (\ref{mulate}), thus potentially serving as an experimental signature.

In view of the above, a natural question is how far into the life of the quark-gluon plasma one could trust our early-time damping results (\ref{atemp}), (\ref{pradtemp}), (\ref{eradtemp}) and (\ref{muearly}). As discussed at the end of the previous section, the validity of our approximations (\ref{smalltemp}) and (\ref{ultrarel}) implies the temporal bounds (\ref{tultrarel}) and (\ref{ttemp}), of which the latter is the most restrictive. We thus know that we are limited to times much smaller than
\begin{equation}\label{ttemp2}
t_{\mbox{\scriptsize temp}}\equiv \frac{1}{\pi T \sqrt{v_m^2-v_0^2}}
\simeq \frac{10}{\pi T}.
\end{equation}
Given that 1/$\pi T$ (for $T_{QCD}\simeq 250$ MeV)
corresponds to $0.25~\mbox{fm}/c$ under the ``obvious'' and
$0.33~\mbox{fm}/c$ under the ``alternative'' prescription of
\cite{gubsercompare}, we see that our limiting time (\ref{ttemp2}) will be only of
 a few $\mbox{fm}/c$. This is an order of magnitude better than the time at which the numerical integration in \cite{dragtime} broke down (for the optimal scenario of a nonrelativistic quark), but is still, at best, of order the lifetime of the quark-gluon plasma, meaning that our results would only be justified in the early phase of the latter's evolution.

 For times of order $t_{\mbox{\scriptsize temp}}$, we certainly expect the solution (\ref{xcorrected1}) to be modified by the appearance of $T^8$ and higher corrections neglected in the string equation of motion (\ref{smalltempngeom}). As the string continues to evolve, we expect a dynamical worldsheet horizon (located at $z=z_h$ in the remote past) to move down toward smaller values of $z$, and then (if the quark is undisturbed) move back up towards the spacetime horizon \cite{dragtime}. When the wavefront on the string crosses the worldsheet horizon, one would have to worry about the choice of boundary conditions there. Throughout this paper, we have employed the same condition as in \cite{mikhailov,lorentzdirac,damping}, which describes waves in the gluonic field that are required to propagate outward from the quark. This captures the causal setup that is most natural for a quark in vacuum, and is therefore also appropriate as long as we consider only small perturbations about (\ref{mikhsol}), as we have done here given that our goal from the outset was to find the first correction due to the thermal medium. When the evolution progresses further, however, it is to be expected that, as the medium is disturbed by the quark, it radiates gluonic waves back towards the source. Needless to say, determining and implementing the precise boundary condition in a fully dynamical setup is bound to be a challenging problem, which would seem to require in particular knowledge of the exact solution of the nonlinear thermal equation (\ref{ngeomtemp}). The net result of the additional thermal effects described in the present paragraph should be to yield a damping coefficient $\mu(t)$ that smoothly interpolates between the early-time result (\ref{muearly}) and the late-time form (\ref{mulate}).

 Beyond the limitation to early times, it should be borne in mind that the setup in this paper, as well as in \cite{hkkky,gubser,ct}, involves an infinite static thermal medium, whereas the real-world quark-gluon plasma is of finite extent, and is exploding and cooling rapidly. Starting with \cite{expandinginitial}, the literature contains various examples of gravity duals for expanding plasmas (for reviews, see, e.g., \cite{expandingrev}), but given the reliance of our approach on the exact zero-temperature embedding (\ref{mikhsol}) derived in \cite{mikhailov} by exploiting the symmetries of the pure AdS background, at present there seems to be unfortunately little hope of being able to extract from them \emph{analytic} results for the early-time thermal damping rate (see, however, \cite{expandingdamping} for  related results).

 Another issue is the choice of initial conditions for the quark. We wish to model a situation where the quark is created within the plasma and is only starting to feel its effects. To account for this, on the gravity side we have been considering a string that is disturbed only in a region close to its endpoint, with the disturbance thereafter traveling deeper into the bulk, but not having had time yet to get close to the black hole horizon. The cleanest scenario involves a string that is originally at rest, and whose endpoint is then accelerated via an external force for some short period of time, after which it is released. This is precisely the situation that was examined numerically in \cite{dragtime}. The period of external forcing can be viewed as modeling the formation time for our energetic quark within the plasma, and it merely serves to set some initial conditions. Both from general physical principles and from the numerical results of \cite{dragtime}, we know that, if this period is short enough, the quark will behave as in vacuum, and so the corresponding evolution of the string will be essentially as in pure AdS. The question we have been addressing in this paper is what happens after the quark is released and slowly begins to feel the effect of the thermal medium.

  Another possibility that might come to mind would be to imagine a quark traveling at constant velocity in vacuum, which then suddenly encounters the thermal medium. A similar setup was explored in \cite{dominguez,hk}. In this case, in the gravity description we would start with a vertical string translating uniformly in pure AdS and then in some fashion `switch on' the $h-1$ metric deformation leading us to the black hole geometry. However, in such a situation,  the portion of the string that first feels the thermal effects would clearly be the one closest to the spacetime horizon at $z=z_h$, where our condition (\ref{smalltemp}) is violated. It is easy to see that the vertical string traveling at constant velocity is in fact a solution of the full thermal equation of motion (\ref{ngeomtemp}), so that is not an issue. The problem, known already from \cite{hkkky,gubser}, is that this string worldsheet becomes spacelike above the critical radial position $z_v\equiv (1-v^2)^{1/4}z_h$, and is therefore unphysical. So, in this scenario where the quark is suddenly introduced into the plasma, there will necessarily be a wave  running \emph{downward} along the string to ensure that a physical boundary condition is achieved at the emerging worldsheet horizon. This is the same type of effect that we argued above to be important to eventually lead to the late-time/stationary rate (\ref{elossgubser}), and for the same reason, it is beyond the reach of our approach. In MSYM language, the initial quark in this scenario has its full vacuum dressing (i.e., the gluonic cloud surrounding it is undeformed out to arbitrarily large distances), and so does not closely mimic the experimental situation.

 Returning to the scenario where the string is initially static and is then forced over a finite period of time, we should emphasize that our use of the vacuum configuration as a starting point \emph{does not} restrict our analysis to situations where the string remains nearly vertical (which would in turn require the forcing not to be too violent), because Mikhailov's solution (\ref{mikhsol}) is valid for arbitrary trajectories of the quark. The vertical string profile is recovered in the case where the quark moves at constant velocity, but an exact description is equally available in situations where the quark is forced (even violently), such as uniform proper acceleration (where the solution was found independently in \cite{xiao,ppz} but was later shown \cite{brownian} to be equivalent to (\ref{mikhsol})) or uniform circular motion (where, again, the solution derived by a different route in \cite{liustirring} actually coincides with Mikhailov's).

 Now, in the preceding discussion we have for simplicity been considering the evolution of an isolated quark as it travels through the plasma, but since the quark should be initially created with its corresponding antiquark, we could expect the latter to have some influence on the former. This concern has been addressed in \cite{dragtime}, following \cite{hkkky} (related, more recent, work may be found in \cite{efjt}). The gravity description involves \emph{two} string endpoints on the D7-branes, which separate from one another. As explained in \cite{dragtime}, there are in fact two cases to consider. A quark-antiquark pair in the adjoint representation, such as would be created by a gluon, corresponds to a $\wedge$-shaped string  with its apex located at $z=z_h$, that is initially of zero width and then proceeds to open up as the string endpoints move away from each other. In this case, since the two halves of the string--- corresponding to the quark and antiquark field configurations--- only meet at the black hole horizon, they do not interact in finite time (at large-$N_c$), and we are back to the isolated quark scenario described above (albeit with some differences introduced by  the details of the forcing stage). This is then the most realistic interpretation of the initial conditions naturally associated with our approach.

 The other case refers to a quark-antiquark pair in the singlet representation, such as would be created by a photon, and corresponds to a $\cap$-shaped string that is initially of zero length, with its midpoint moving deeper into the bulk as the two endpoints separate. In this case, it was found numerically in \cite{dragtime} that the evolution is essentially as in vacuum up to the point where the quark and antiquark lose contact by moving beyond their (velocity-dependent) screening length \cite{liuwind,dragqqbar,sonnenschein}, after which their evolution is already controlled by the \emph{late}-time damping (\ref{elossgubser}). Our results (or their generalization to finite external forcing) would then be expected to be relevant only for the initial stage, prior to screening. In this connection we should note that, even though the general vacuum solution (\ref{mikhsol}) is nominally associated with an embedding that has a single endpoint on the D7-branes, and therefore describes an isolated quark, it is in fact also capable of describing at least certain types of $\cap$-shaped profiles, with two endpoints on the D7-branes, such as the example already mentioned two paragraphs above, where the quark and antiquark separate back-to-back with constant proper acceleration. Again, the period of external forcing we have envisioned in our above discussion on initial conditions can be regarded as modeling the stage of formation of the quark, and, in particular, the influence of the antiquark.

Altogether, then, we believe the results obtained in this paper shed some additional light on the nature of early-time energy loss in a scenario that is not too distant from the one that is of phenomenological interest, but of course, additional work is needed to achieve further realism.

\section*{Acknowledgements}

We are grateful to Vadim Kaplunovsky for useful comments, and to Mariano Chernicoff for collaboration in the early stages of this project, for valuable discussions and for helpful comments on the manuscript. The present work was partially supported by Mexico's National Council of Science and Technology (CONACyT) grant 104649.

\end{document}